\documentstyle[11pt,aaspp4,psfig,epsf,flushrt]{article}  
\received{RECEIPT DATE 1994}
\accepted{ACCEPT DATE 1994}
\journalid{VOL}{JOURNAL DATE 1994}
\articleid{START PAGE}{END PAGE}
             \font\sevenrm=cmr7

          \font\sixrm=cmr6       
                     
\def\pr{Phys. Rev.}                             
\def\ssr{Space Sci. Rev.}                       
\def\teq#1{$\, #1\,$}                           

\def\Tbn{\Theta_{\hbox{\sevenrm B}n1}}
\def\Tbntwo{\Theta_{\hbox{\sevenrm B}n2}}
\def\machson{{\cal M}_{\hbox{\sixrm S}}}
\def\machalf{{\cal M}_{\hbox{\sixrm A}}}
\def\machtot{{\cal M}_{\hbox{\sixrm T}}}
\def\dover#1#2{\hbox{${{\displaystyle#1 \vphantom{(} }\over{
   \displaystyle #2 \vphantom{(} }}$}}
\def\horule{\hrule height0.5pt}              
\def\htrule{\hrule height0.7pt}              
\def\vspone{\vphantom{\bigl{)}}}   \def\vsptwo{\vphantom{\Bigl{)}}}
   
    \def\ds{$\;\,$}     
\begin{document}
%
%
\newcommand{\vol}[2]{$\,$\rm #1\rm , #2.}                 
\newcommand{\figureout}[2]{\centerline{}
   \centerline{\psfig{figure=#1,width=5.5in}}   
    \figcaption{#2}\clearpage } 
\newcommand{\figureoutsmall}[2]{\centerline{\psfig{figure=#1,width=5.0in}}
    \figcaption{#2}\clearpage } 
\newcommand{\tableout}[4]{\vskip 0.3truecm \centerline{\rm TABLE #1\rm}
   \vskip 0.2truecm\centerline{\rm #2\rm}   
   \vskip -0.3truecm  \begin{displaymath} #3 \end{displaymath} 
   \noindent \rm #4\rm\vskip 0.1truecm } 
%
%
\title{ACCELERATION OF SOLAR WIND IONS BY NEARBY\\ 
       INTERPLANETARY SHOCKS: COMPARISON OF\\
       MONTE CARLO SIMULATIONS WITH ULYSSES OBSERVATIONS}
   \author{Matthew G. Baring\altaffilmark{1}}
   \affil{Laboratory for High Energy Astrophysics, Code 661, \\
      NASA Goddard Space Flight Center, Greenbelt, MD 20771, U.S.A.\\
      \it baring@lheavx.gsfc.nasa.gov\rm }

   \author{Keith W. Ogilvie}
   \affil{Laboratory for Extraterrestrial Physics, Code 692, \\
      NASA Goddard Space Flight Center, Greenbelt, MD 20771, U.S.A.\\
      \it u2kwo@lepvax.gsfc.nasa.gov\rm}

   \author{Donald C. Ellison}
   \affil{Department of Physics, North Carolina State University,\\
      Box 8202, Raleigh NC 27695, U.S.A.\\
      \it don\_ellison@ncsu.edu\rm}

  \and

   \author{Robert J. Forsyth}
   \affil{The Blackett Laboratory, Imperial College,\\
      Prince Consort Road, London SW7 2BZ, U.K.\\
      \it r.forsyth@ic.ac.uk\rm}
   \altaffiltext{1}{Compton Fellow, Universities Space Research Association} 
   \authoraddr{Laboratory for High Energy Astrophysics, Code 661,
      NASA Goddard Space Flight Center, Greenbelt, MD 20771, U.S.A.}
\slugcomment{To appear in \it The Astrophysical Journal\rm , 
     February 20, 1997 issue.}
%
\clearpage

\begin{abstract}  
Various theoretical techniques have been devised to determine
distribution functions of particles accelerated by the first-order
Fermi mechanism at collisionless astrophysical shocks.  The most
stringent test of these models as descriptors of the phenomenon of
diffusive acceleration is a comparison of the theoretical predictions
with observational data on particle populations.  Such comparisons have
yielded good agreement between observations at the quasi-parallel
portion of the Earth's bow shock and three theoretical approaches,
namely Monte Carlo kinetic simulations, hybrid plasma simulations, and
numerical solution of the diffusion-convection equation.  Testing of
the Monte Carlo method is extended in this paper to the realm of
oblique interplanetary shocks: here observations of proton and
\teq{He^{2+}} distributions made by the SWICS ion mass spectrometer on
Ulysses at nearby interplanetary shocks (less than about 3AU distant
from the sun) are compared with test particle Monte Carlo simulation
predictions of accelerated populations.  The plasma parameters used in
the simulation are obtained from measurements of solar wind particles
and the magnetic field upstream of individual shocks; pick-up ions are
omitted from the simulations, since they appear, for the most part, at
greater heliospheric distances.   Good agreement between downstream
spectral measurements and the simulation predictions are obtained for
two shocks by allowing the parameter \teq{\lambda /r_{\rm g}}, the
ratio of the mean-free scattering length to the ionic gyroradius, to
vary in an optimization of the fit to the data; generally \teq{\lambda
/r_{\rm g}\lesssim 5}, corresponding to the case of strong scattering.
Simultaneous \teq{H^+} and \teq{He^{2+}} data, presented only for the
April 7, 1991 shock event, indicate that the acceleration process is
roughly independent of the mass or charge of the species.   This
naturally arises if all particles interact elastically with a massive
background, as occurs in collisionless ``scattering'' off a background
magnetic field, and is a patent property of the Monte Carlo technique
since it assumes elastic and quasi-isotropic scattering of particles in
the local plasma frame.
\end{abstract}  
\clearpage 
\section{INTRODUCTION}

Particle acceleration at collisionless shocks is believed to be a common
phenomenon in space plasmas belonging to a diversity of environments,
ranging from the inner heliosphere to the central regions of distant
galaxies (e.g. Blandford and Eichler 1987).  Evidence to support the
belief that such a mechanism can efficiently produce non-thermal
particles includes direct measurements of accelerated populations in
various energy ranges at the Earth's bow shock (e.g. Scholer et al.
1980; Ipavich, Scholer and Gloeckler 1981; M\"obius et al. 1987; Gosling
et al. 1989) and interplanetary shocks (for the pre-Ulysses era see, for
example, Sarris and Van Allen 1974; Gosling et al. 1981; Decker, Pesses
and Krimigis, 1981; Kennel et al. 1984; Tan et al. 1988), and indirect
evidence provided by the observation of non-thermal electromagnetic
radiation in distant cosmic sources such as supernova remnants and
exotic galaxies and the very existence of cosmic rays.  The motivations
for developing theories of shock acceleration are therefore obvious, and
several popular approaches have emerged over the last twenty years.  One
of these is the Monte Carlo technique of Ellison and Jones (e.g. see
Jones and Ellison, 1991, and references therein) which nicely describes
the injection and acceleration of particles from thermal energies, at
the same time minimizing the complexities of the underlying microphysics
while retaining the essential ingredients of the Fermi acceleration
mechanism.

While theoretical work can provide considerable understanding of the
diffusive shock acceleration process, the test of its appropriateness
to physical environments can only be made by comparison of model
predictions with observational data.  This has been done very
successfully (Ellison, M\"obius and Paschmann 1990; see also Ellison
and M\"obius 1987) in the case of the quasi-parallel portion of the
Earth's bow shock, comparing spectral predictions of the aforementioned
Monte Carlo simulation with proton, He$^{2+}$ and C, N and O ion
particle distributions obtained by the AMPTE experiment.  Using the
solar wind temperatures of the different ionic species as input for the
simulation upstream of the bow shock, downstream particle distributions
were obtained from the simulation and compared with the AMPTE data: the
agreement was impressive.  An important conclusion of this comparison
was that the data implied that the dynamic effects of the accelerated
particles are crucial to the determination of the shock structure.
This feature is a consequence of the quasi-parallel bow shock being a
fairly strong shock (i.e., Mach number about 7) and a comparatively
efficient accelerator ($\gtrsim 15\%$ of the solar wind energy flux
goes into superthermal particles; see Ellison, M\"obius and Paschmann
1990).  Since this pioneering work, successful comparisons of other
theoretical techniques with data from the Earth's bow shock have been
performed.  These include the hybrid plasma simulations of Trattner and
Scholer (1991, 1993) and of Giacalone and coworkers (described in
Giacalone, Burgess, Schwartz and Ellison 1992, 1993), which yielded
quite reasonable agreement with both the bow shock data and the Monte
Carlo technique (Ellison et al. 1993).  Very recently, Kang and Jones
(1995) have obtained solutions to the convection-diffusion equation for
particle acceleration at parallel shocks that generate particle
distributions similar to the bow shock observations and both the Monte
Carlo and hybrid simulations.

\clearpage

It is natural to extend such comparisons to other shock environments.
The Ulysses spacecraft has provided the opportunity for an application
of this work  to interplanetary (IP) shocks, and the high quality and
comparative abundance of the spectral data (e.g. see Ogilvie, et al.
1993; Gloeckler, et al. 1995; Baring et al. 1995) obtained by the SWICS
instrument aboard Ulysses enhances the significance of such tests of
theoretical models.  This paper presents  results from our program of
fitting the Ulysses observations at oblique interplanetary shocks with
Monte Carlo simulation output.  This has become possible following the
generalization of the Monte Carlo technique to treat oblique shocks
(Baring, Ellison and Jones 1993, 1994) and to examine cases of strong
scattering turbulence (Baring, Ellison and Jones 1995; Ellison, Baring
and Jones 1995). Shorter presentations of these comparisons between
theory and experiment are enunciated in Baring et al. (1995) and
Baring, Ogilvie and Ellison (1995).  In this paper, our simulations are
confined to test particle implementations for simplicity, which is
quite appropriate for the weak IP shocks considered here (discussed in
Section~4.2).  Note that this is not the first attempt to fit
interplanetary shock data with Monte Carlo simulation results:  Ellison
and Eichler (1984) fit  International Sun-Earth Explorer (ISEE) 3
proton data (taken from Gosling et al. 1981) of the IP shock detected
on 27th August 1978 with considerable success (see also Ellison 1983
for simultaneous \teq{He^{2+}} data fits).  However, their simulations
were restricted to the case of truly parallel shocks, while the
observed shock was only quasi-parallel. Therefore, having simulations
of oblique shocks now at our disposal, we are in a position to model IP
shocks more realistically than Ellison and Eichler (1984) and assess
how the strength of particle scattering influences the acceleration
process.  We note also that Jones and Kang (1995) and Kang and Jones
(1996) have very recently applied their convection-diffusion equation
technique to the IP shock \teq{H^+} data presented in Baring et al.
(1995), again with considerable success in their theory/observation
comparison.

In this paper we compare the SWICS measurements of proton and He$^{2+}$
distributions downstream of interplanetary shocks observed at 2.7--3.2AU
from the sun with the simulation output, using upstream measurements of
the solar wind quantities to determine the shock parameters for the
model. Using a single free parameter \teq{\lambda /r_{\rm g}}, the ratio
of the mean-free path for scattering to the ionic gyroradius, to
describe the strength of the scattering turbulence, excellent fits of
the theory to the observations are obtained for two of our chosen shocks
when strong scattering is present, i.e. when \teq{\lambda /r_{\rm
g}\lesssim 10}.  We regard this as significant confirmatory evidence for
the existence of strong turbulence in IP shocks, a result that is borne
out in magnetic field data (e.g. Tsurutani, Smith and Jones 1983 for 
ISEE 3 observations; Vi\~nas, Goldstein and Acu\~na 1984 for Voyager
spacecraft measurements; Balogh et al. 1993 and Section~5.2 here for
Ulysses magnetometer results). Furthermore, the simultaneous
observations of \teq{H^+} and \teq{He^{2+}} spectra are remarkably
similar (except for normalization) when plotted on a velocity scale,
indicating that the acceleration process treats particles of a given
speed almost identically despite differences in mass or charge. Such a
situation is expected if all particles interact elastically with a
massive background; the Monte Carlo technique naturally reproduces this
definitive property as a consequence of its assumption of elastic
scattering of particles in the local plasma frame.  Our attention is
confined here to shocks less than 3.5AU distant from the sun,
encountered by Ulysses in the first half of 1991.   Few pick-up ions, in
particular protons, are evident at heliospheric distances less than
around 3.2AU.  Interstellar pick-up ions, which can participate in the
acceleration process, are not incorporated in the simulation results
presented here. The reader is referred to Gloeckler et al. (1994, 1995)
for interesting Ulysses observations of IP shocks with strong signals
from pick-up hydrogen, \teq{He^+} and \teq{He^{2+}}.  Before detailing
our fitting procedure and results in Section~4, we describe briefly the
SWICS instrument in Section~2 and the Monte Carlo simulation (see
Section~3). The paper concludes with a discussion of the differences
between our IP shocks and the quasi-perpendicular portions of the
Earth's bow shock, possible signatures of pick-up protons in one of our
candidate shocks, and concurrent Ulysses magnetometer data.

\section{THE SWICS SPECTROMETER ON ULYSSES}

The Solar Wind Ion Composition Spectrometer (SWICS) on the Ulysses spacecraft
measures the intensity and distribution functions of solar wind and
suprathermal ions over a range of energy per charge from \teq{0.6} to \teq{60}
keV/e in \teq{64} logarithmically spaced steps with energy resolution
\teq{\Delta E/E\sim 0.04}.  The energy-per-charge analysis is combined with
post-acceleration through a potential difference of \teq{23}kV, and followed by
time-of-flight and energy measurement to identify ionic mass and charge states. 
Because of the double and triple coincidences used, the background levels are
very low, typically less than one count per hundred hours.  This permits
measurements of very low flux levels.  The instrument uses a curved multi-slit
collimator, which covers \teq{57^\circ\times 4^\circ} in angle, fixed to the
spacecraft with the spacecraft spin axis passing through one end.  Thus the
instrument sweeps out a cone of half-angle \teq{57^\circ} as it rotates.  The
cone is centered approximately in the solar wind direction, and is sampled once
per spacecraft revolution, which is 12 seconds in duration.  The entire 
velocity range of observed ionic distribution functions is covered 
sequentially in 64 of these samples, i.e. once every 13 minutes; hence any
variability in the ion populations is undetectable on shorter timescales.  In
order to restrict the counting rate of the ``start'' channel plate in the
time-of-flight system, the SWICS data processing unit has a mode where the
voltage sweep applied to the electrostatic analyser is reversed if the counting
rate exceeds a preset number.  Since the voltage sweep tracks particles from
high to low energy, this safety feature (Gliem et al. 1988) prohibits
measurement of the ion distributions much below the channel with peak count
rate.  This mode was in use early in the Ulysses mission, indeed for the data
presented here, but its use was subsequently abandoned, having been deemed
unnecessary for instrumental security.

Data used in this manner implicitly assumes that the detected ions are
samples of a distribution that is isotropic within this cone, an
expedient approximation that is extremely difficult to relax.  Careful
consideration of particle anisotropies is therefore essential to more
refined analysis of the experimental data.  For example, the solar wind
beam is always completely sampled by a small portion of the collimator,
whereas ions accelerated to well above 1000 km/sec are quasi-isotropic
and therefore occupy a larger solid angle than does the rotating
collimator.  Between these cases are suprathermal ions, whose inferred
distribution depends somewhat on the precise degree of anisotropy in the
plasma, an unmeasured and therefore assumption-dependent quantity. 
While such intricacies of data analysis  do not affect the qualitative
nature of the scientific conclusions presented (this matter will be
briefly addressed in Section~4.2), it must be noted that it is the
suprathermal (or injected) ions, which have the largest uncertainties
due to unknown anisotropies, that are the most influential in the shock
acceleration process. For more complete details of the SWICS instrument,
see Gloeckler et al. (1992).  This study uses measurements of protons
and doubly-charged helium only; data on \teq{He^+} pick-up ions observed
at greater heliospheric distances is presented in Gloeckler et al.
(1994).  Also, note that data at energies higher than \teq{60}keV have
been obtained by the HI-SCALE experiment on Ulysses (described in
Lanzarotti et al. 1992).  Use of such data was not possible at this
stage, and is deferred to future work.


\section{MONTE CARLO SIMULATIONS OF SHOCK ACCELERATION}

The simulation technique (described in detail in Ellison, Jones and
Eichler 1981; Jones and Ellison 1991; Baring, Ellison and Jones 1993;
Ellison, Baring, and Jones 1996) is a kinematic model, closely
following Bell's (1978) approach to diffusive acceleration; it
essentially solves a Boltzmann equation for particle transport
involving a collision operator.  Particles are injected upstream and
allowed to convect into the shock (i.e. mimicking the solar wind),
colliding with postulated scattering centers (presumably magnetic
irregularities) along the way.  As they diffuse between the upstream
and downstream regions, they continually gain energy (for a simulation
example, see Fig.~3 of Baring, Ellison and Jones 1994), sampling the
velocity differential across the shock; this is the principle of the
Fermi acceleration mechanism.  An important property of the model is
that it treats thermal particles like accelerated ones, making no
distinction between them. Hence, as the accelerated particles start off
as thermal ones, this technique automatically injects particles from
the thermal population into the acceleration process, thus
circumventing the injection problem that plagues some models of shock
acceleration.  One valuable consequence of this unified treatment for
thermal and non-thermal particles is that modification of the shock
hydrodynamics by the accelerated population can easily be incorporated
to forge a fully self-consistent model of acceleration and shock
structure.  Such non-linear hydrodynamics were included in the
modelling of the quasi-parallel portion of the Earth's bow shock
(Ellison, M\"obius and Paschmann 1990), but is not important for the
interplanetary shocks examined in this paper: this is evident from the
spectral data presented in Section~4.2 below.  Shock modification is
therefore not included in the present modelling.  However, we note that
it may prove necessary to include such non-linear effects when
performing fits over a much broader range of energies than is the case
here, for example when including HI-SCALE data.

The scattering is assumed to be elastic and to isotropize particle
momenta in the local fluid frame, i.e. to mimic large-angle collisions
with field turbulence that is anchored in this frame.  Such collisions
might well be expected in view of the quite turbulent fields observed in
heliospheric shock environments (e.g. Hoppe et al. 1981; Tsurutani,
Smith and Jones 1983; Balogh et al. 1993).  In the simulation, particles
are allowed to travel between scatterings for a time that is
exponentially distributed about the time \teq{t_c=\lambda/v}, where
\teq{v} is the particle speed in the fluid frame and \teq{\lambda} is
the scattering mean free path in the fluid frame, which is taken to be a
power-law function of momentum \teq{p}:
\begin{equation}
   \lambda\; =\; \lambda_0\; \biggl(\dover{r_{\rm g}}{r_{\rm g1}}
   \biggr)^{\alpha}\; \propto\; p^{\alpha}\quad .
  \label{eq:mfp}
\end{equation}
Here \teq{r_{\rm g}=pc/(QeB)} is the gyroradius of a particle with
momentum \teq{p}, charge \teq{Qe}, in a magnetic field of strength
\teq{B} (\teq{c} is the speed of light and \teq{e} is the electronic
charge), and \teq{r_{\rm g1}=m_p u_{1x}/(eB_1)} is the gyroradius of a
proton with the far upstream speed \teq{u_{1x}} (i.e. the component of
the upstream flow velocity normal to the shock plane in the rest frame
of the shock) in the upstream magnetic field, \teq{B_1}.
Hereafter, \teq{x} denotes distance measured along the
shock normal, and subscripts 1(2) will indicate upstream (downstream). 
Here \teq{\lambda_0} is set proportional to the gyroradius
\teq{r_{\rm g1}=mu_{1x}/(eB_1)} of (low energy) protons of speed
\teq{u_{1x}}: the ratio \teq{\lambda_0/r_{\rm g1}}  is an input parameter
that is of crucial importance to this study.  For simplicity,
\teq{\lambda_0/r_{\rm g1}} is taken to be the same both upstream and
downstream of the shock, a subjective choice that is largely immaterial
to the nature of the results presented below. 

This simple prescription is perhaps surprisingly appropriate to shocked
plasma environments.  Bow shock observations (Ellison, M\"obius, and
Paschmann 1990) indicate that \teq{1/2 < \alpha < 3/2}, while \teq{1/2 <
\alpha < 4/5} can be deduced from ions accelerated in solar particle
events (Mason, Gloeckler and Hovestadt 1983), and \teq{\alpha\sim 1/2}
is obtained from turbulence in the interplanetary magnetic field
(Moussas et al. 1992).  Plasma simulations
(Giacalone, Burgess, and Schwartz 1992) suggest a mean free path obeying
Eq.~(\ref{eq:mfp}) with \teq{\alpha \sim 2/3}.  Clearly a variety of
microphysics can be incorporated in specifying \teq{\lambda} and both
pitch-angle diffusion (small-angle scattering) and large-angle
scattering can be modelled by the simulation.  Therefore the
attractiveness of this approach is that the overall nature of the
diffusive acceleration process can be modelled without the burden of CPU
intensive computations of the underlying physical plasma mechanisms. 
All results presented here are for \teq{\alpha =1}, a choice that nicely
guarantees that the ratio \teq{\lambda /r_{\rm g}} is independent of the
particle momentum.

The assumption of elasticity of scattering in the local fluid frame is
expected to be accurate except perhaps at shocks with very low
Alfv\'enic Mach numbers, where the flow speed does not far exceed the
Alfv\'en velocity and the scattering centers (i.e. Alfv\'en waves) move
in the fluid frame with significant speeds.  In the shocks studied
here, as is typically the case for interplanetary shock systems, the
magnetic field is dynamically important, with Alfv\'enic Mach numbers
in the range of \teq{2.4}--\teq{4}, so that to some degree the elastic
scattering assumption is violated.  Nevertheless, for the shock
parameters relevant to this study, the contribution to particle
acceleration resulting from motion of the scattering centers in the
local fluid frame is expected to be overshadowed by the first-order
acceleration process that is modelled in this paper, and which
principally determines the overall shape of model particle
distributions.  We expect that the major properties of the acceleration
mechanism are therefore well-described by elasticity of scattering, and
that the deviations from this in IP shock environments will incur only
effects of secondary importance.  Hence, in the spirit of this
incipient comparison of theory and experiment, we adopt the elastic
scattering approximation, and defer examination of the higher order
effects that are associated with its violation to future work.

In applications of the Monte Carlo simulation to oblique shocks,
particle transport is followed in the de Hoffmann-Teller (HT) frame  (de
Hoffmann and Teller 1950), where  the drift electric field \teq{{\bf E}
= ({\bf u}/c) \times {\bf B}} is transformed to zero, and the fluid flow
is everywhere parallel to the magnetic field.  This approach guarantees
that shock drift acceleration is automatically included.   Also, since
our technique does not require detailed computation of the turbulent
field structure, it is much faster than hybrid or full plasma codes, 
and thereby achieves much larger dynamic ranges  of momenta for particle
distributions than is possible for the plasma simulations.  This renders
the Monte Carlo approach ideal for spectral comparisons with
observational data.  There are two variations of the simulation that are
employed here: (i) one in which a guiding-center approximation is made
(Baring, Ellison and Jones 1994), where the details of the particles'
gyromotions are ignored and they conserve their adiabatic moment
\teq{p^2 (1-\mu^2)/2B} (the so-called \it adiabatic approximation\rm )
when interacting with the shock; and (ii) where the exact gyrohelix
particle trajectories (e.g. see Ellison, Baring and Jones 1995, 1996)
are computed without imposition of the adiabatic approximation in
particle-shock interactions.  Results from both variations are employed
in this paper.  Another simplification made in the simulation is the use
of a single parameter \teq{\lambda_0/r_{\rm g1}} to describe the strength of
the scattering turbulence (i.e. cross-field diffusion; see Baring,
Ellison and Jones, 1995 for details).  This parameter plays a crucial
role in determining the efficiency of injection of particles into the 
non-thermal tail, as is discussed in Section~4.

Our model is implemented in this paper as a test particle simulation,
where the accelerated particles make only a small contribution to the
dynamics of the fluid flow.  This is appropriate for modelling
interplanetary shocks, which are of low sonic Mach number and therefore
comparatively inefficient at accelerating particles (as discussed in
Section~4.2).  For a full discussion of the non-linear effects
associated with high Mach number shocks, see Ellison, Baring, and Jones
(1996). The usual hydrodynamic quantities of the downstream fluid are
therefore simply determined from those in the upstream region by the
full MHD Rankine-Hugoniot relations (e.g. Decker 1988).  In the IP
shock applications here the  magnetic field is high enough to be
dynamically important.  Hence, all characteristics of the shocks depend
only on  the shock compression ratio \teq{r=u_{1x}/u_{2x}}, the sonic
Mach number \teq{\machson\approx\lbrack m_p u_{1x}^2/(\gamma k
T_1)\rbrack^{1/2}}, where \teq{T_1=T_{p1}+T_{e1}} is the total upstream
temperature, \teq{\gamma =5/3} is the adiabatic index, and the
Alfv\'enic Mach number \teq{\machalf \approx \lbrack 4\pi n_pm_p
u_{1x}^2/B^2\rbrack^{1/2}}.  Simulation output principally includes
particle distributions in any desired frame of reference at any
position upstream or downstream of the shock.


\section{COMPARISON OF MONTE CARLO SPECTRA WITH SWICS DATA}

Before comparing in detail the Monte Carlo particle distributions with
the SWICS measurements downstream of our three case-study IP shocks, it
is appropriate first to indicate what input parameters for the
simulations were derived from various pieces of observational data.  The
three shocks chosen were observed early in the Ulysses mission,
specifically on day 97 of 1991 (April 7; denoted 91097), day 118 of 1991
(April 28; denoted 91118), and day 147 of 1991 (May 27; denoted 91147),
when Ulysses was in the ecliptic before its encounter with Jupiter. 
Motivation for our choices included availability of high quality spectra
(i.e. with good statistics) in these shocks, and \it relatively \rm
simple shock structure and spectra during this phase of the mission. 
These shocks appear to be representative of the shocks encountered by
Ulysses in the first half of 1991.  The prominence of accelerated
particles in the environs of many shocks other than those studied
here suggests that the basic physical conclusions drawn in this paper
can be extended to a host of other IP shocks detected during this
early phase of the Ulysses mission.

Of the three shocks in our case study, two (91097 and 91118) are freely
propagating (travelling) interplanetary shocks and the other (91147) is
a member of a co-rotating shock pair.  \it Co-rotating interaction
regions \rm (CIRs e.g. see Hundhausen 1973 for a description),
frequently observed by Ulysses (e.g. Gosling et al. 1993, 1995), result
from the evolution of regions where solar wind regions of higher speed
impinge upon lower speed wind that emanates from the equatorial region
of the sun.   At the heliocentric distances considered here, the
collision results in a forward and reverse shock pair on either side of
an interface; the shocks propagate away from the interface, which itself
moves away from the sun at a speed greater than that of the solar wind. 
We have selected the forward shock on day 91147 only for our study.

\subsection{Observational Parameters and Simulation Input}

Following the approach that Ellison, M\"obius and Paschmann (1990) used
when comparing Monte Carlo spectra with observations at the
quasi-parallel portion of the Earth's bow shock, we use here
measurements of various solar wind quantities upstream of three
interplanetary shocks to set the input parameters for the Monte Carlo
simulation runs.  The simulation then makes predictions of downstream
spectra, which are then compared with  SWICS data downstream of these IP
shocks to ascertain whether consistency between experiment and theory
can be achieved.  

The various input quantities used in the Monte Carlo runs are tabulated
in Table~1 for the three selected IP shocks, together with their
corresponding heliocentric distances.  All the quantities in the upper
zone except for the shock velocities \teq{u_s} were directly determined
from measurements by various instruments on Ulysses; \teq{u_s}, the ion
temperatures and the shock Mach numbers are inferred from the measured
quantities. Specifically, \teq{u_s} for shocks 91097 and 91147 are taken 
from the published listing of Burton et. al. (1992), which cites a shock
speed of 2 km/sec for 91118.  This erroneous result, probably
typographical, can be corrected by inferring the value of \teq{u_s}
based on the other measurable quantities and the total Mach number of
1.97 cited by Burton et. al. (1992) and Balogh et al. (1992).  The
result is an estimate of 165 km/sec, which implicitly has at least a
10\% uncertainty since the Mach number can be no more accurate than the
least certain measurable, which is the field obliquity 
\teq{\Theta_{\hbox{\sevenrm B}1}}.  Unpublished data on plasma speeds
combined with the measured obliquity were used to verify the consistency
of this estimate.

\placetable{Table1}

The values listed in Table~1 for the thermal velocities \teq{v_{th}} 
were taken from Gaussian fits to spectral data less than an hour
upstream of the respective shocks.  The peak of the phase space
distribution was fit with a form proportional to \teq{\exp
[-v^2/v^2_{th}]}.  Upstream densities \teq{n_{H^+}} and
\teq{n_{He^{2+}}} are quoted for similar timescales.  The upstream
magnetic field values  \teq{B_1} are averages on timescales of about a
minute, and their accuracy is limited by temporal/spatial fluctuations,
which are discussed in Section~5.  As a consistency check on the
tabulated values, we used the Monte Carlo code to generate solutions of
the Rankine-Hugoniot relations for the field compression ratio
\teq{B_2/B_1} from the input solar wind quantities, and these agreed to
within about 8\% of the ratios published by experimental teams (Balogh
et al. 1995, Hoang et al. 1995).  The solar wind electron temperatures
for the 91097 and 91147 shocks are published in Hoang et al. (1995),
where they indicate large uncertainties (\teq{\sim 30\%}).  No such
listing for the 91118 event was available, so we used the best fit
radial dependence temperature at 2.85 AU obtained from Phillips et al.
(1995), representing a length scale of the order of 0.25AU.  This choice
is subject to large uncertainty since the electron temperature varies on
short timescales.  With the exception of the shock obliquity,
uncertainties for the measured solar wind quantities largely reflect
limits set by fluctuations in these quantities and the choice of
temporal binning.  Determination of  \teq{\Theta_{\hbox{\sevenrm B}1}}
is intrinsically problematic because it is difficult to establish the
plane of the shock in a turbulent environment with measurements made
only along the spacecraft trajectory.

Clearly the uncertainties in most shock parameters are of the order of
10\%, which must be taken into account in the following theory-data
comparison.  Uncertainties are not quoted in Table~1 for the derived
quantities, i.e. ion temperatures and Mach numbers.  As indicated, the
values of \teq{\machtot} listed in Table~1 for 91097 and 91147 are
derived using the measured values in the upper zone, whereas the value
of 2.07 for the 91118 shock was used to estimate the shock velocity. 
Simple formulae for the sonic and Alfv\'enic Mach numbers that are
scaled to suit typical IP shock parameters can be quickly written down:
\begin{equation}
   \machson\;\approx\; 8.52\,\dover{u_{100}}{\sqrt{T_{p4}+T_{e4}}}\quad ,\quad
   \machalf\;\approx\; 4.58\,\dover{u_{100}}{B_{-5}}\,\sqrt{n_{H^+}}\quad ,
  \label{eq:machnos}
\end{equation}
where \teq{u_{100}} is the shock speed \teq{u_s} in units of 100 km/sec,
\teq{T_{p4}} and \teq{T_{e4}} are the proton and electron temperatures
in units of \teq{10^4}K, \teq{B_{-5}} is the upstream field strength in
units of \teq{10^{-5}}Gauss, and \teq{n_{H^+}} is the proton density in
\teq{\hbox{cm}^{-3}}, which is approximately equal to the electron
density \teq{n_e} (an 8\% difference arises due to the alpha particle
abundance in the solar wind). The comparable values of these two Mach
numbers for the IP shocks studied here indicate that the fields are
dynamically important; their proximity to unity suggests that the shocks
are weak and therefore are relatively inefficient particle accelerators
(e.g. compared to the Earth's bow shock), a feature that is borne out in
their spectral data.  Note that we neglected to explicitly include the
effects of \teq{He^{2+}} on the dynamics of the 91097 event.  While
alpha particles comprise about 4\% if the solar wind population, and
therefore are about 16\% abundant by mass, they also comprise roughly
the same percentage of the thermal energy (i.e. 16\% ; a coincidence
specific to 91097). Hence, dynamically they contribute about the same
fraction of the ram pressure (i.e. the numerator of \teq{\machson^2})
and the thermal pressure (denominator) and hence have marginal
influence on the value of \teq{\machson}. Helium data are considered
only for the April 7, 1991 shock since they were of almost comparable
quality to the proton data; the non-thermal tails of the He spectra for
the 91118 and 91147 events had insufficient statistical quality for the
purposes of this analysis.

\subsection{The Spectral Comparison}

The comparisons of the Monte Carlo model results with the IP shock data
are performed in the spacecraft frame, using boost speeds between the
shock and spacecraft frames in the range of around 360--430 km/sec,
depending on the shock.  Such speeds were deduced from the speeds
associated with the thermal peaks upstream and downstream of each of
the shocks, and were consistent with estimates (subject to large
uncertainties) determined from the shock parameters listed in Burton
et al. (1992).  A multiplicative factor is applied to the Monte Carlo
upstream proton distributions to render them commensurate with the
solar wind density. The fitting of the observed spectral data with the
simulation output is then essentially achieved by adjusting a \it
single \rm free model parameter \teq{\lambda_0/r_{\rm g1}} by visual
inspection.  This parameter, whose physical significance is discussed
shortly, is a scaling parameter for the simulation, being the ratio of
the mean free path to the gyroradius of particles with speed
\teq{u_{1x}} (see Section 3).  The scientific conclusions of this paper
are amply defined via this simple visual procedure.

\placefigure{fig:91118gc}

Figure~\ref{fig:91118gc} depicts the data/theory comparison for the IP
shock on April 28, 1991 (day 91118) using the guiding center (GC)
version of the simulation.  The experimental data are taken over a
period of a few hours on the downstream side of the shock where the
distribution was relatively steady in time; throughout this paper the
uncertainties on the data points denote one standard deviation.  A
similar comparison is made in Fig.~\ref{fig:91118orb} using the
gyrohelix (GH) version of the code.  The two fits are quite similar
with the gyrohelix version requiring a somewhat lower value of
\teq{\lambda_0/r_{\rm g1}} to approximate the data. Note that this
period of data collection is chosen to be quite long in order to
achieve good statistics.  The base instrumental spectral time
resolution of 13 minutes corresponds to the order of 400 gyroperiods
\teq{T_g} for protons behind the shock (where \teq{B\sim 5\times
10^{-5}}Gauss),  where \teq{T_g\equiv\omega_c^{-1}=m_pc/(eB)\approx
10.4/B_{-5}} sec.  Here \teq{\omega_c} is the gyrofrequency  and
\teq{B_{-5}} is the field in units of \teq{10^{-5}}Gauss.  Hence the
much longer integration times used in the spectral comparisons of this
Section generally represent several thousand gyroperiods.  The
properties of the shock itself are deduced from the much more frequent
measurements of the magnetic field.

The SWICS particle distribution data used here and throughout this paper
are phase space densities \teq{f(v)}, for which \teq{v^2f(v)\, dv}
represents a particle density; these are essentially the count rate
(i.e. detected flux) divided by velocity to the fourth power.    As the
velocities are in units of km/sec, the distribution functions depicted
here integrate over velocity to yield densities in units of particles
per cubic kilometre.  Since data are taken in the spacecraft frame, the
thermal peak exhibits a large positive offset of a few hundred km/sec
relative to the frame of the shock, reflecting the solar wind speed in
the ecliptic. The counts are dead-time corrected, but do not include
complete corrections for instrumental geometry and response.  Such
corrections are not yet fully implemented, but are expected to produce
at most about a factor of a 2--3 difference to the distribution. Thus
these corrections are small over the dynamic range of seven orders of
magnitude for the distributions in the figures of this section, and it
is evident that their inclusion would not significantly alter the
physical conclusions deduced from the comparisons presented.

Figures~\ref{fig:91118gc} and~\ref{fig:91118orb} can be used to
estimate the contribution of the accelerated population to the dynamics
of the 91118 shock.  The phase space densities \teq{f(v)} used here can
be multiplied by a factor of \teq{v^4} to yield energy density
distributions.  This weighting should be considered in the shock rest
frame, where treatment of the dynamics is simplest.  In this frame, for
the IP shocks considered here, the downstream thermal particles
typically have speeds around \teq{150}km/sec (i.e. the shock speeds
listed in Table~1).  Hence in transforming data from the spacecraft to
the shock rest frames, clearly a ratio of around 10 in speeds is
encompassed in the shock frame by the distributions in
Figs.~\ref{fig:91118gc} and~\ref{fig:91118orb}.  It follows that the
high speed particles have at most a few percent of the energy density
contained in the thermal population, a result that extends to the other
two shocks.  The accelerated populations detected by the SWICS
instrument are therefore dynamically unimportant, thereby justifying
the test particle implementation in this paper.

\placefigure{fig:91118orb}

The agreement between experiment and theory in Figs.~\ref{fig:91118gc}
and~\ref{fig:91118orb} is impressive, with only small differences in the
downstream temperature (for the guiding center implementation of the
code) and shape of the suprathermal portion of the distributions being
apparent.  The fits are of comparable quality for the two types of
particle convection employed, namely (i) exact gyrohelix determination
and (ii) use of the guiding center approximation with the added
assumption of adiabatic moment conservation.
Furthermore, the data strongly constrains the possible values of
\teq{\lambda_0/r_{\rm g1}} (typically \teq{\pm 0.1}).
While these two versions of the simulation produce
somewhat different estimates of the parameter \teq{\lambda_0/r_{\rm g1}} due
to their differing prescriptions for the way particles cross the shock,
the physical conclusions drawn from these values are similar.  It
appears that exact gyrohelix computations are slightly less efficient at
injecting particles from the thermal population in highly oblique
shocks.  The physical importance of this parameter follows from its
crucial role in determining the extent of cross-field diffusion. If
\teq{\kappa_\parallel} and \teq{\kappa_\perp} are the diffusion
coefficients respectively parallel and perpendicular to the average
magnetic field, then the two can be related using the standard kinetic
theory result (e.g. Axford 1965; Forman, Jokipii and Owens 1974) for a
mean free path \teq{\lambda} and particle gyroradius \teq{r_{\rm g}} in the
local fluid frame:
\begin{equation}
   \kappa_\perp\; =\;\dover{\kappa_\parallel}{1 +(\lambda/r_{\rm g} )^2 }\quad .
  \label{eq:kappas}
\end{equation}
Non-zero values of \teq{\kappa_\perp} (i.e. non-infinite \teq{\lambda
/r_{\rm g}}) amount to the presence of cross-field diffusion.  This
identity is tantamount to shifting a particle by approximately one
gyroradius in {\it any direction} during scatterings with the turbulent
magnetic field.  Jokipii (1987) observed that there is no general
consensus as to the form of this relationship, so that the choice of
the ratio is somewhat subjective, being restricted only by the
requirement that a particle move of the order of a gyroradius in a
scattering.  Clearly then, large values of \teq{\lambda /r_{\rm g}}
(\teq{\equiv\lambda_0/r_{\rm g1}} in this application) represent weak
cross-field diffusion, where the field lines are almost laminar, while
small values (\teq{\lambda /r_{\rm g}\ll 1}) define the regime of Bohm
(isotropic) diffusion, which may be attained in strongly turbulent
fields.

Using an analysis of particle acceleration times, Ellison, Baring and
Jones (1995) demonstrated that the Monte Carlo technique accurately
simulates the kinetic theory result in equation~\ref{eq:kappas}.
Besides confirming Jokipii's (1987) observation that shocks of higher
obliquity are \it faster \rm accelerators, the analysis of Ellison,
Baring and Jones (1995) indicated that the efficiency of injecting
particles into the Fermi acceleration mechanism is a strongly \it
decreasing \rm function of shock obliquity and the strength of the
scattering parameter \teq{\lambda_0/r_{\rm g1}} (see also Baring,
Ellison and Jones 1993).  This sensitivity is a crucial feature of the
acceleration process that compels the range of \teq{\lambda_0/r_{\rm
g1}} in the fit to be narrow.  The origin of this dependence can be
simply understood.  In the de-Hoffmann Teller frame, the diffusion
length scale of low energy (i.e.  thermal) particles along the field is
roughly given by \teq{\lambda_0}, while their diffusion length normal
to the field is represented by the gyroradius \teq{r_{\rm g1}}.  In
oblique shocks, specifically those with \teq{\Tbn\gtrsim 30^\circ},
convective return of thermal ions to the shock from the downstream side
is inhibited so that the injection into the Fermi acceleration
mechanism is suppressed (Baring, Ellison and Jones 1993, 1994).  From
simple considerations of geometry, this is clearly circumvented by
cross-field diffusion when \teq{r_{\rm g1}/\lambda_0\gtrsim\cot\Tbntwo}
(for \teq{\Tbntwo} being the downstream field obliquity) and particle
return to the shock is enabled (Baring, Ellison and Jones 1993;
Ellison, Baring and Jones 1995). Hence it is expected that the
efficiency of injection and acceleration will scale inversely with the
parameter \teq{(\lambda_0/r_{\rm g1})/(1+\tan\Tbntwo)}.

Clearly then, when \teq{\lambda_0/r_{\rm g1}\gg 1} and cross-field
diffusion is small, the accelerated population is strongly suppressed
in highly oblique shocks, contrary to the observational data.  The
light solid histogram in Fig.~\ref{fig:91118gc} is the predicted
simulation in the limit of \teq{\lambda_0/r_{\rm g1}\to\infty}, starkly
illustrating the lack of acceleration from thermal energies.  Therefore
in the fits in Figs.~\ref{fig:91118gc} and~\ref{fig:91118orb}, where
the high velocity portions of the histograms shift up and down rapidly
relative to the thermal peaks with only modest variations in
\teq{\lambda_0/r_{\rm g1}}, only small values of \teq{\lambda_0/r_{\rm
g1}} are allowed.  It can be simply deduced that, within the context of
this Monte Carlo model, the significant acceleration efficiency implies
that strong scattering is operating in this traveling interplanetary
shock.  A similar conclusion is reached in the subsequent presentation
for other IP shocks, which are also efficient accelerators, and is
suggestive of turbulent magnetic fields in interplanetary shock
environments. Since the efficiency is extremely  sensitive to the shock
obliquity, the most uncertain of the observables, improved measurements
will refine estimates of \teq{\lambda_0/r_{\rm g1}}, but will not alter
the implication of strong scattering acting in interplanetary shocks.
Note that hereafter, \teq{\lambda_0/r_{\rm g1}} and \teq{\lambda/r_{\rm
g}} will be used interchangeably due to their identity, which is the
result of assuming \teq{\alpha =1} in the scattering formula in
equation~(\ref{eq:mfp}).

\placefigure{fig:91097}


Figure~\ref{fig:91097} displays the spectral comparison for the IP shock
on April 7, 1991 (day 91097) using the guiding center (GC) version of
the simulation.  The SWICS data are again taken over a period of a few
hours (i.e. thousands of gyroperiods) on the downstream side of the
shock, but this time, the data permitted presentation of the
\teq{He^{2+}} phase space distribution also. The relative normalizations
of the upstream distributions for protons and \teq{He^{2+}} were deduced
from the abundance ratio implied by Table~1,  and are not free
parameters in the model.  The fit is clearly impressive for the proton
spectrum, and is equally encouraging for the alpha particles considering
that just  \it one \rm parameter (\teq{\lambda /r_{\rm g}}) is used to
reproduce both the spectral shape \it and \rm the relative populations
of both accelerated ion species.  Again, the relatively low value of 
\teq{\lambda /r_{\rm g} = 3.7} implies strong scattering.

At first sight, it seems remarkable that just one choice of \teq{\lambda
/r_{\rm g}} can fit the the two species so nicely.  However, this is
consistent with the observed spectra being essentially independent of
mass and charge, i.e. depending only on velocity kinematics.  Since the
inferred upstream \teq{He^{2+}} temperature is about four times that of
the protons, the input for the simulation is roughly independent of mass
and charge.  If, in addition, all particles interact nearly elastically
with a massive background, namely magnetic field turbulence
``anchored'' in the fluid frame, the acceleration process, and therefore
the spectral shape and injection efficiency, will depend only on
particle velocities.  Since the Monte Carlo simulation makes just this
elastic assumption (i.e. in Eq.~[\ref{eq:mfp}]), it is not surprising
that we can accurately model the observations in this test-particle
regime.  Given such dependence on the kinematics, the relative
normalizations of the \teq{H^+} and \teq{He^{2+}} data downstream are
largely determined by their relative abundances upstream.  The results
in Fig.~\ref{fig:91097} indicate the high accuracy of such a scattering
assumption in traveling interplanetary shocks when interstellar pick-up
ions are unimportant.  

Another noteworthy feature of this comparison is that this is the first
time that the upstream abundance ratio {\it of thermal ions} has been
used to set relative population levels in a theory/data comparison of
acceleration at shocks somewhere in the heliosphere. Due to the
inability of the 3D-Plasma Instrument on the AMPTE/IRM spacecraft to
measure elements heavier than hydrogen, the study of the parallel
portion of the Earth's bow shock by Ellison, M\"obius and Paschmann
(1990) did not have upstream measurements at thermal energies of
elements heavier than hydrogen (downstream observations of non-thermal
alpha particles and heavier elements were made above 10 keV/Q by the
SULEICA instrument), and so could not determine relative abundances in
the local solar wind.  It should be noted that the values of
\teq{\lambda /r_{\rm g}} obtained in the fits presented here differ
somewhat from previous expositions of this data/theory comparison,
notably in Baring et al. (1995) and Baring, Ogilvie and Ellison
(1995).  This difference arises because more refined estimates of the
magnetic field and solar wind densities have been used here than in the
earlier works, with finite electron temperatures being included,
thus revising the input parameters for the shock simulation.
The shock strength is somewhat weaker than before, so that estimates
for \teq{\lambda /r_{\rm g}} are now lower.  We note that recently with
the aid of shock parameters from Baring et al. (1995),  Kang and Jones
(1995) and Kang and Jones (1996) have obtained, comparable fits to the
proton data for the 91097 and 91118 shocks using their
convection-diffusion equation model of the Fermi acceleration mechanism.

\placefigure{fig:91147}

As a contrast to the above data comparisons, it is instructive to study
a shock that was encountered at a later stage of Ulysses' passage away
from the sun.   Figure~\ref{fig:91147} displays the comparison for the
IP shock on May 27, 1991 (day 91147), again using the guiding center
(GC) version of the simulation, and proton data accumulated over several
hours. The choice of \teq{\lambda /r_{\rm g}} is solely influenced by the fit
to the high velocity portion (1000--1800 km/sec) of the data, and again
suggests strong scattering.  However, the value of \teq{\lambda /r_{\rm g}} is
somewhat higher in this shock than in the previous two, due to its lower
obliquity.  The inaccuracy of the comparison in the suprathermal region
contrasts strongly with the previous fits, and is probably indicative of
the presence of another proton population (i.e. interstellar pick-up
ions),  as is discussed below, rather than representing the failure of
the Fermi acceleration mechanism.  The presence of such pick-up ions,
with speeds ranging between zero and twice the solar wind speed, may
account for the discrepancy, since the simulation deficiency in
Fig.~\ref{fig:91147} is in precisely this velocity range.  The present
implementation of the Monte Carlo simulation models injection and
acceleration of particles drawn only directly from the thermal
population, however, it can be easily adapted to describe additional
pre-accelerated populations.     

To summarize, in two of our three case studies excellent fits have been 
obtained, though some small discrepancies exist at thermal and
suprathermal energies.  The slight deficiencies of the simulation
predictions relative to the data in
Figs.~\ref{fig:91118gc}--\ref{fig:91097} (particularly for the guiding
center implementations of the simulation) could be due to subtleties in
the microphysics that are omitted from the simulation, such as the
presence of a cross-shock potential that would modify the acceleration
efficiency, or deviations from the scattering law in
equation~(\ref{eq:mfp}).  Furthermore, there may be some suprathermal
population present at a level smaller than in Fig.~\ref{fig:91147} that
may distort the  distributions from the simple simulation predictions. 
In addition, the omission from the simulation of non-linear
modifications of the shock hydrodynamics by the accelerated population
(generally small for IP shocks) might lead to a slightly incorrect
estimation of the downstream temperature.  Such non-linear effects
smooth the shock velocity profile upstream (decelerating the flow) and
increase the overall compression ratio, thereby modifying the width of
the downstream thermal distribution (e.g.  Ellison, Baring and Jones
1996).  It must be remembered that the fits attempted here are confined
to relatively low particle speeds, and may be degraded or modified
somewhat by a future consideration of data at higher energies, such as
that obtained by HI-SCALE, where the test-particle Fermi acceleration
distributions assume well-known (e.g.  see Blandford and Eichler 1987;
Jones and Ellison 1991) power-laws.

\section{DISCUSSION}

Perhaps the most remarkable aspect of the observations presented here is
simply that low Mach number, highly oblique, interplanetary shocks
accelerate significant fractions of thermal solar wind protons and alpha
particles to non-thermal energies, and do so with comparable efficiencies.
The smooth transition from the thermal portion of the distribution to 
a non-thermal tail seen in Figs.~\ref{fig:91118gc}--\ref{fig:91097}, and,
to a lesser extent, \ref{fig:91147}, is a distinctive signature of the
direct injection and acceleration of thermal particles, given that no
such transition is seen less than an hour upstream of these shocks. 
This close association of thermal and non-thermal ions in the
neighbourhood of these quasi-perpendicular IP shocks is decidedly different
behavior from that observed at highly oblique portions of the Earth's
bow shock, and also from plasma simulation results, which show no
significant acceleration of thermal ions at all in shocks of high obliquity. 

At quasi-perpendicular parts of the bow shock, reflected field-aligned
beams are typically seen propagating back from the shock, and these
beams have a much lower \teq{He^{2+}}/proton ratio than the ambient
solar wind (e.g. Ipavich et al.  1988; Fuselier and Thomsen 1992). These
beams are believed to originate when particles specularly reflect off
the shock (see Thomsen 1985 for a review), with protons reflecting more
easily than \teq{He^{2+}}. Downstream from the bow shock, the helium
remains at much lower relative density compared to the protons (Fuselier
and Schmidt 1994), in contrast to what we see here downstream from IP
shocks. The radically different behavior of highly oblique regions of
the Earth's bow shock and the IP shocks considered here is probably the
result of the very different relative length scales involved.  Both
travelling IP shocks and those associated with CIRs have radii of
curvature that are much larger than that of the earth's bow shock, at
least for large distances from the sun.  The diffusion length of
suprathermal particles at the bow shock is comparable to the bow shock
radius, and reflected or accelerated particles that are produced near
the tangent point of the solar wind magnetic field with the shock, where
the field has very little turbulence, are quickly swept away and do not
have the chance to generate local turbulence.  An IP shock, on the other
hand, is much larger compared to diffusion length scales so that
particles generated at quasi-perpendicular shock environments can remain
in the system long enough to maintain much larger levels of turbulence.

An idea of the scale lengths associated with these shock systems can be
gained from the gyroradii of the detected particles.  At
non-relativistic velocities \teq{v}, a particle of mass number \teq{A}
and charge state \teq{Q} has a gyroradius of \teq{r_{\rm g}=(A/Q)\, m_p
vc/(eB)}, which can be translated into units of relevance to the plasma
environment considered here:
\begin{equation}
   r_{\rm g}(\hbox{km})\; =\; \dover{A}{Q}\,\dover{10.43}{B_{-5}}\, 
   v_{\hbox{\sevenrm km/sec}}\; =\; \dover{A}{Q}\,\dover{4567}{B_{-5}}\,
   \sqrt{E_{\hbox{\sevenrm keV}}}\quad ,
  \label{eq:rgforms}
\end{equation}
where \teq{B_{-5}} is the field strength expressed in units of
\teq{10^{-5}}Gauss, and \teq{E_{\hbox{\sevenrm keV}}} is the particle
energy in keV.  Clearly, since the IP shock velocities are of the order
of 100--300 km/sec, thermal protons have gyroradii around \teq{10^3}km,
and even the most energetic ions detected by SWICS have gyroradii (and
therefore mean free paths) much less than the typically huge IP shock
scale (\teq{>10^7}km).  This behaviour differs strongly from that at the
Earth's bow shock, which has particles accelerated to comparable
energies but a much smaller radius  curvature, of the order of
\teq{10^4}km.   An important byproduct of these scale length estimates
is that the use of a planar shock model in this study is both expedient
and justifiable for interplanetary shock applications.  The angle
between the shock normal and the magnetic field, \teq{\Tbn}, is constant
for much longer distances along the faces of such shocks than typical
particle mean free paths.  As the efficiency of shock acceleration is a
strong function of \teq{\Tbn} (e.g. Baring, Ellison and Jones 1993;
Ellison, Baring and Jones 1995, 1996), interplanetary shocks are very
suitable for comparison with theory, since the observations at a given
point with given \teq{\Tbn} are not contaminated by particles
accelerated at locales of the shock where \teq{\Tbn} is different.

In regards to full plasma simulations, to our knowledge, all published
1-and 2-dimensional plasma simulation results for shocks with $\Tbn
\gtrsim 60^{\circ}$ fail to show significant acceleration of thermal
ions (e.g. Liewer, Goldstein, and Omidi 1993; Liewer, Rath, and
Goldstein 1995; Kucharek and Scholer 1995).  We believe this to be a
consequence of an unphysical suppression of cross-field diffusion in
simulations with field geometries of less than three dimensions
(Jokipii, K\'ota, and Giacalone 1993; Jokipii and Jones 1996).  Since
cross-field diffusion is essential for allowing downstream particles to
recross the shock, as noted above, its suppression will eliminate
particle injection and possibly also subsequent self-generation of
field turbulence.  The kinematic Monte Carlo technique simply
incorporates a prescription of such particle diffusion via the
parameter \teq{\lambda /r_{\rm g}}, and hence is inherently
three-dimensional in its description of particle motions.

\subsection{Pick-up Ions?}

The excellence of the data fits to the 91097 and 91118 shocks, and the 
comparatively poor fit to the 91147 data, could be considered consistent
results if an extra population of seed particles for the acceleration
mechanism were present in the 91147 event, thereby circumventing the
deficiency seen in Fig.~\ref{fig:91147}.   Interstellar pick-up ions,
which are neutral when they enter the heliosphere, are ionized by the
solar UV flux or by charge exchange with the solar wind (e.g. Gloeckler
and Geiss 1996). They naturally present themselves as candidates for
such a population, especially since they have a velocity distribution
ranging between zero and roughly twice the solar wind speed (e.g.
Vasyliunas and Siscoe, 1976), i.e. exactly where the Monte Carlo model
underpredicts the accelerated population for the 91147 event.   This
population is particularly interesting since the acceleration efficiency
of suprathermal ions exceeds that of the thermal particles. The density
of pick-up ions in the heliosphere is not uniform due to the
gravitational focusing of the ion streams in the solar neighbourhood; in
fact, the degree of focusing differs for protons and \teq{He^+}
(Vasyliunas and Siscoe, 1976)  so that the radial variations of the
densities of these pick-up species are not the same.  Such density
gradients are also influenced by variations in the ionization fractions. 
Helium atoms, whose ionization potential is high, penetrate as close as
0.5AU, whereas hydrogen gas is highly ionized by the time it is less
than a few AU from the sun.  Hence the density gradient for \teq{H^+} is
much greater in the 2AU--5AU range than for singly-ionized helium. Note
that while \teq{He^+} pick-up ions are much more prevalent than 
\teq{He^{2+}} pick-up ions, \teq{He^{2+}} is the dominant charge state
of helium in the solar wind plasma.

Fortunately, due to measurements by SWICS, as presented by Gloeckler and
Geiss (1996), we have a strong indication of the variations in pick-up
ion backgrounds at times when the spectra presented in this study were
obtained. They indicate that for \teq{He^+}, a slow decline in count
rate was observed as Ulysses made its way from 1AU to beyond Jupiter
(from 2 counts to 0.8 counts per 12 second instrumental sweep), whereas
the flux ratio of \teq{H^+/He^+} increased by an order of magnitude
between 3AU and 5AU.   Consequently, it can be inferred that the proton
pick-up density increased dramatically during this period of data
collection, and specifically, that an increase of a factor of \teq{\sim
3-5} occurred between the 91097 and 91147 events.  The pick-up ion count
rate in the 91147 shock was actually a significant fraction 
of the proton count rate at comparable suprathermal energies. This implies
that the data/theory comparison presented here can comfortably
accommodate the hypothesis that there is a very small contribution of
pick-up protons for the 91097 shock observed at \teq{\sim 2.7}AU, but
that this contribution has risen dramatically by day 147 in 1991,
corresponding to \teq{\sim 3.2}AU.  We therefore contend that pick-up
ions play an important role in the acceleration process in the 91147
event, and that our omission of them from the Monte Carlo model is
primarily responsible for the poorness of the fit.

\subsection{$\lambda /r_{\rm g}$ and Field Turbulence}

The question of plausibility of the values for \teq{\lambda/r_{\rm g}}
obtained by our fits of the spectral data naturally arises.  Precedence
for such low \teq{\lambda/r_{\rm g}} exists in the literature.  Forman, Jokipii
and Owens (1974) determined the ratio
\teq{\kappa_{\perp}/\kappa_{\parallel}} of components of the diffusion
tensor along and orthogonal to the field using quasi-linear diffusion
theory applied to measurements of the interplanetary magnetic field
power spectrum made by the Mariner 4 spacecraft at locations near earth
and also near Mars.  They computed a ratio of \teq{0.07}, which
corresponds to \teq{\lambda/r_{\rm g}\sim 4} using the kinetic theory formula
in equation~(\ref{eq:kappas}).  Similar estimates are obtained at much
larger solar distances (\teq{\sim 15}--20 AU) using somewhat different 
analyses of magnetic field data obtained by Voyager 2, to compute
\teq{\kappa_{\perp}} (see Valdes-Galicia, Quenby and Moussas 1992), and
also field data acquired by Pioneer 11 to determine
\teq{\kappa_{\parallel}} (see Moussas et al. 1992).  While these results
were obtained using measurements made near the ecliptic plane, it is
anticipated that similar estimates will result near the heliospheric
poles from information recently acquired by the Ulysses mission.  The
values for \teq{\lambda /r_{\rm g}} determined from the spectral comparisons
here are therefore not extraordinary by heliospheric standards.  Yet the
diffusion coefficients computed by the analyses of Forman, Jokipii and
Owens (1974), Valdes-Galicia, Quenby and Moussas (1992), and Moussas et
al. (1992) are all in the range of about \teq{3\times 10^{20}}cm$^2$
sec$^{-1}$--\teq{3\times 10^{22}}cm$^2$ sec$^{-1}$, implying a mean free
path of around 1--100 AU for particles of speeds \teq{v\sim 1000}km/sec 
(\teq{\lambda\sim 3\kappa /v}).  These far exceed the estimate of a few
times \teq{10^4}km (\teq{\sim 3\times 10^{-4}}AU) obtainable from
equation~(\ref{eq:rgforms}) for similar \teq{v} for the IP shocks of studied
here. This dramatic difference is very probably not a conflict, but
rather reflects a huge disparity in spatial scale of diffusion in the
general interplanetary medium as opposed to diffusion in the environs of
interplanetary shocks.  It must be emphasized that the spectral
estimates of \teq{\lambda/r_{\rm g}} determined here are very sensitive to the
shock obliquity \teq{\Tbn}, a quantity that is difficult to measure
accurately in space plasma shocks.

\begin{figure}  \epsfysize=20.5cm
   \hbox to\hsize{\hfill \hbox{\epsfbox{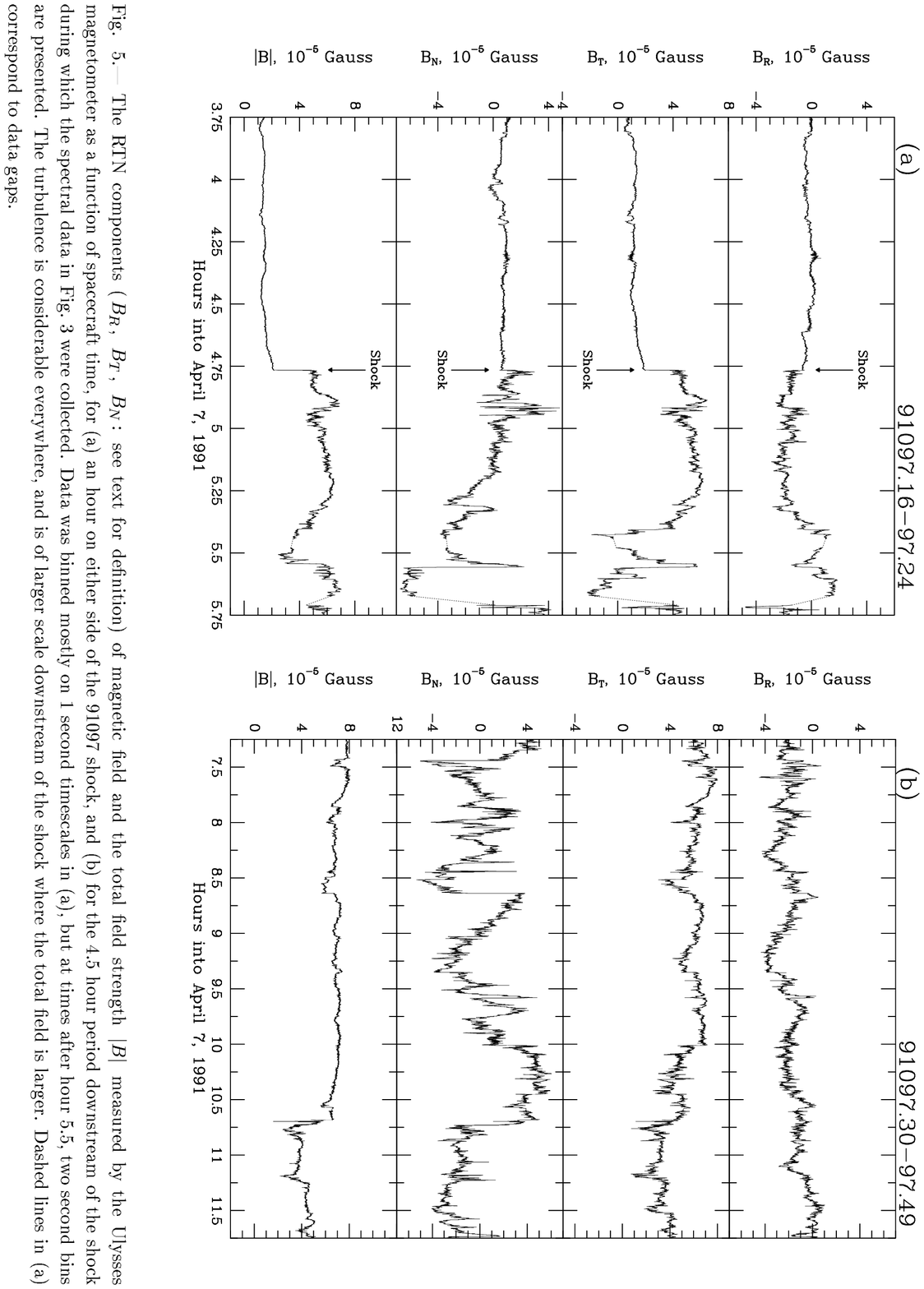}} \hfill} 
      \label{fig:Bcomp}
\end{figure}

\placefigure{fig:Bcomp}

The diffusion length scale associated with interplanetary shocks can be
examined by consideration of the magnetic field turbulence in their
environs.   Fig.~\ref{fig:Bcomp} displays the magnetic field data
measured by the Ulysses magnetometer in the temporal vicinity of the
91097 IP shock on 1 sec (up to the 5.5 hour mark) and 2 sec (more than
5.5 hours into April 7, 1991) timescales.  The orthogonal RTN
components \teq{B_R}, \teq{B_T}, \teq{B_N}, of magnetic field and the
total field strength  \teq{\vert B\vert} comprise the data streams. Here
the \teq{R}-direction points radially away from the sun, the
\teq{N}-axis is perpendicular to the \teq{R}-axis and lying in the plane
formed by the radial vector to the spacecraft and the solar rotation
axis, and the \teq{T}-direction completes the orthogonal coordinates
such that R,T, and N form a right-handed system. In
Fig.~\ref{fig:Bcomp}a the field quantities are presented for the period
of an hour on either side of the 91097 shock to illustrate  properties
coupled to the shock on smaller scale lengths.  The field turbulence
appears much larger on the downstream side of the shock (the right hand
side of the plot, at later times), but this is somewhat deceptive since
there the field has been enhanced due to the shock compression (see the
bottom panel).  The level of turbulence relative to the mean ambient
field is physically a more meaningful quantity, and will be addressed
shortly.  Also in Fig.~\ref{fig:Bcomp}a there is evidence of sharp
discontinuities in field components downstream. These are not shocks
since they are not accompanied by steep density gradients, yet they are
representative of spatial/temporal structural features  generally seen
in the turbulent environs of IP shocks.  Field data streams for the
other candidate shocks look very similar to Fig.~\ref{fig:Bcomp}.  In
Fig.~\ref{fig:Bcomp}b the field data are displayed for the 4.5 hour
period downstream of the shock during which the spectral data in
Fig.~\ref{fig:91097} were collected, revealing no significantly
different behaviour from the downstream section in 
Fig.~\ref{fig:Bcomp}a.

\placefigure{fig:turbulence}

A more illuminating presentation of the turbulent magnetic field is
given in Fig.~\ref{fig:turbulence}, which exhibits the field \it
turbulence measure \rm \teq{\vert \delta\hbox{\bf B}\vert/\vert\hbox{\bf
B}\vert} derived from the measured data in Fig.~\ref{fig:Bcomp}a.  The
turbulence measure, which can formally be related to the magnetic
Reynolds number \teq{R_{\hbox{\sixrm M}}\sim 2\vert\delta\hbox{\bf
B}\vert^2 /(\machalf\vert\hbox{\bf B}\vert )^2} (e.g. see Boyd and
Sanderson 1969, p.~126), was obtained by comparing vector components of
the field in successive time bins (labelled \teq{i} and \teq{i+1})
according to 
\begin{equation}
   \machalf\,\sqrt{\dover{R_{\hbox{\sixrm M}}}{2}}\;\sim\;
   \dover{\vert\delta\hbox{\bf B}\vert}{\vert\hbox{\bf B}\vert}\; =\;
   \dover{1}{\vert\hbox{\bf B}\vert}\,\sqrt{\delta B_R^2 +
   \delta B_T^2+\delta B_N^2} \; =\;\dover{\vert\hbox{\bf B}_{i+1}-
   \hbox{\bf B}_i\vert}{\vert\hbox{\bf B}_i\vert}    \label{eq:turb} 
\end{equation} 
on 1s, 10s and 100s binning
timescales.  While the choice of turbulence measure is subjective, this
definition is informative, and virtually coincides with definition of
the normalized variance that is frequently used in the space plasma
literature: see Forsyth et al. (1996) for variances during 
out-of-the-ecliptic passages of Ulysses, which are generally taken
as averages over longer terms rather than successive data bins.  The
remarkable feature of Fig.~\ref{fig:turbulence} is the similarity of the
turbulence measure on both sides of the shock: the shock enhances the
field and the turbulence by similar factors so that the relative level
of turbulence remains more or less the same.  It is unclear whether
or not such a compressive enhancement is caused by shock-generated
turbulence.  It should be noted that any differences in turbulence
measure between the upstream and downstream regions would argue for
different values for \teq{\lambda/r_{\rm g}} either side of the shock. 
However, for the Monte Carlo technique, the key acceleration properties
of spectral shape and injection efficiency depend mostly on the value of
\teq{\lambda/r_{\rm g}} in the downstream region, so that only this value need
be specified. The similarity of the turbulence measures throughout the
shock transition further indicates that the added complication of treating
such non-uniformities in \teq{\lambda/r_{\rm g}} is unneccesary.

The turbulence measure averages around 5\% on second timescales, which
is probably around the border of the quasi-linear regime, and increases
to an average of 16\% on 100 second timescales, reflecting a non-uniform
power spectrum.  This latter level of turbulence is quite substantial
and is suggestive that quasi-linear theory may not be applicable on
these larger timescales.  Note that \teq{\vert \delta\hbox{\bf
B}\vert/\vert\hbox{\bf B}\vert >1} at the shock on the 1 and 10 second
timescales.  The binning is truncated at the onset of data gaps in
Fig.~\ref{fig:Bcomp}a.  Without a more sophisticated analysis
(e.g. see Forman, Jokipii and Owens 1974), it is
difficult to conclude whether or not these turbulence measures are
closely consistent with the spectrally-determined estimates of
\teq{\lambda/r_{\rm g}}.  However, it is possible to rule out extremities. The
Bohm diffusion limit \teq{\lambda\sim r_{\rm g}} is expected to arise when
\teq{\vert \delta\hbox{\bf B}\vert/\vert\hbox{\bf B}\vert \gtrsim 0.5},
which is not reached in this data.  Yet the levels of turbulence are
high enough that the plasma is probably much closer to the Bohm
diffusion limit than to a \teq{\lambda/r_{\rm g}\sim 30} state.  Since the
proton gyroperiod is \teq{T_g=m_pc/(eB)\approx 10.4/B_{-5}} sec, then
protons behind the shock (where \teq{B\sim 5\times 10^{-5}}Gauss) sample
fluctuation timescales of the order of a few seconds.  This provides a
rough guide to the level of turbulence expected to dominate the particle
diffusion.

\subsection{Fermi Acceleration}

Another consistency check concerns the acceleration timescale: are the
IP shocks old enough to permit the diffusive Fermi mechanism to
accelerate ions to the energies observed?  Numerous theoretical
treatments of diffusive acceleration timescales exist in the literature,
and we refer the reader to Jones and Ellison (1991) for a review.  For
the Monte Carlo technique used here, where the mean free path is
proportional to the ion rigidity, an explicit form for  the acceleration
time to a given energy \teq{E} is presented in equation~(4) of Ellison,
Baring and Jones (1995), and order of magnitude estimates can be gleaned
from Fig.~3 of that work.  Specifically, the acceleration time
\teq{\tau_{\rm a}} is approximately proportional to \teq{E}, and yields
values of the order of minutes for energization to 10 keV, the typical
upper range of SWICS observations, and timescales of the order of a few
hours for the MeV energies detected by HI-SCALE.  Both these timescales
are considerably shorter than the multi-day duration of passage of our
candidate IP shocks through the heliosphere, thereby causing no temporal
consistency problems for the Fermi mechanism.

It is instructive to remark upon the upstream IP spectral structure and
the simulation predictions.  The SWICS data upstream of the three
candidate shocks are dominated by the beam-like distribution of the solar
wind that is thermally-broadened, containing minimal evidence of
suprathermal ions.  Hence, the measured Upstream densities of
\teq{n_{H^+}} and \teq{n_{He^{2+}}} (for event 91097) are 
representative of the solar wind properties, containing virtually no
pollution from non-thermal ions.  Yet, there is marginal evidence in the
SWICS data for particles seen upstream at low flux levels but with
speeds considerably in excess of the solar wind speed.  No continuity of
the distribution function is apparent, and it could even be described as
consisting of two components of disparate velocities. The upstream
non-thermal ``component'' appears only near the shock. Evidence for such
a non-thermal population that wanes with distance from the shock has
been seen upstream of  interplanetary shocks by previous instruments. 
Gosling et al. (1984) display distributions that diminish in intensity
upstream of IP shocks encountered by ISEE 1 and 2, and that exhibit
upstream non-thermal distributions that are flatter than their
downstream counterparts.  Tan et al. (1989) present ISEE 3 proton data
that impressively exhibits a non-thermal distribution that flattens at
greater distances upstream of an IP shock passage.  

Such properties are consistent with the theory of Fermi acceleration. 
Lee's (1982) analytic modelling of ion acceleration and wave excitation
at the Earth's bow shock predicts that low energy suprathermal ions
disappear at short distances upstream of the shock, with a depletion
between the solar wind and high speed particles that grows both in size
and range of energy as the distance upstream increases.  Our Monte Carlo
model displays virtually identical properties (Baring, Ellison and Jones
1994; Ellison, Baring and Jones 1996), which are a direct consequence of
low energy suprathermal particles having shorter mean paths than ions of
higher energies, and therefore having greater difficulty diffusing
against the convective power of the solar wind to a given distance
upstream of the shock.  The ISEE data add weight to the hypothesis that
Fermi acceleration is acting at interplanetary shocks.

\section{CONCLUSION}

The spectral comparison between SWICS measurements of ion distributions
downstream of three interplanetary shocks observed at 2.7--3.2AU from
the sun, and our kinematic Monte Carlo model for diffusive shock
acceleration, provides impressive fits of the data for two events
(detected on days 91097 and 91118) with theory by effectively varying
just a single parameter, \teq{\lambda/r_{\rm g}}.  The low values
(\teq{\lesssim 5}) of this parameter indicate strong cross-field
diffusion (not far from the Bohm diffusion limit) is operating in these
systems, a property that is commensurate with magnetic field turbulence
in the general interplanetary medium, and is consistent with
self-generated turbulence predicted in diverse shock models.  
Furthermore, the turbulent magnetic field data presented for the 91097
shock appears, without sophisticated analysis, to be concordant with low
values of \teq{\lambda/r_{\rm g}} deduced from the spectral
considerations.  It is anticipated that the inference of strong diffusion
normal to the mean ambient magnetic field direction  applies to many of
the IP shocks encountered by Ulysses during the first half of 1991,
since they exhibit similar evidence of prominent downstream non-thermal
ion populations.

The Monte Carlo model simulates these complex shock systems in an
elegantly simple but elucidating manner, using the measured upstream
solar wind quantities as input for the theoretical simulation.  It
successfully models proton and He$^{2+}$ distributions in the 91097
event, both in spectral shape and normalization, using the \it same \rm
value of \teq{\lambda/r_{\rm g}}.  This is because the acceleration
process at IP shocks seems dependent only on the velocity of the
species, not its mass or charge.  Such a kinematic property is  not
expected if reflection at the shock layer is an important part of the
injection process (e.g. at the quasi-perpendicular portion of the
Earth's bow shock).  However, it is an inherent assumption of the Monte
Carlo technique in the test-particle acceleration regime that is
appropriate for the  generally weak IP shocks.  The theory makes clear,
concise and eminently testable predictions about the fall-off of the
accelerated population upstream of shocks; evidence for this phenomenon
occurring at interplanetary shocks already exists in the literature.  In
the case of the May 27, 1991 shock, the model underpredicts the proton
data in the 500--1000km/sec range.  We propose that this discrepancy
might be rectified if interstellar pick-up protons, which were observed
at significant levels during late May 1991 and at subsequent stages of the
Ulysses mission, are included
in the simulation as an extra population seeding the acceleration
process.  Anticipated developments of the present work include extension
of the comparison to other shocks, and analysis of the field data for
the three shocks considered here, using quasi-linear theory, to obtain
alternative estimates for the ratio \teq{\lambda /r_{\rm g}}.

\acknowledgments
We thank Drs. Tom Jones and Len Burlaga for helpful comments following
careful reading of the manuscript.  We are especially grateful Dr.
George Gloeckler, Prof. Johannes Geiss, and the many individuals at the
University of Maryland, the University of Bern, the Technische
Universit\"at Braunschweig,  the Max-Planck-Institut f\"ur Aeronomie,
and Goddard Space Flight Center who have contributed to the SWICS
experiment.  We have benefited from discussions with Dr. Daniel
Berdichevsky of Hughes/STX, who has been responsible for production of
the data.  We also thank Dr. Frank Jones, whose encouragement led to
our theory/experiment collaboration and the instigation of this
project.  This work was supported by NASA/JPL contract 955460, the
Swiss NSF, the Bundesminister f\"ur  Forschung and Technologie of
Germany, and the NASA Space Physics Theory Program.  DCE thanks  the
Service d'Astrophysique, CEN-Saclay, Observatoire de Paris--Meudon, and
CNET/CETP (Issy-les-Moulineaux)  for hospitality during much of the
period in which work for this paper was completed.

\clearpage

\tableout{1}{OBSERVED SOLAR WIND QUANTITIES: INPUT FOR MONTE CARLO SIMULATIONS}
{                                                                           
\vbox{\offinterlineskip
\htrule
\halign{\strut\ds\hfil$#$\hfil\ds  &  \ds\vrule#\ds  &  \ds\hfil$#$\hfil\ds  &
    \ds\hfil$#$\hfil\ds  &  \ds\hfil$#$\hfil\ds  &  \ds\vrule#\ds  &
    \ds\hfil$#$\hfil\ds  \cr
\vsptwo  && \hbox{April 7, 1991} & \hbox{April 28, 1991}
    & \hbox{May 27, 1991} && \cr
\vspone && 2.7\, \hbox{AU} & 2.85\, \hbox{AU} & 3.15\, \hbox{AU} && \cr 
\vsptwo  && \hbox{(91097)} & \hbox{(91118)} & \hbox{(91147)} && \hbox{Notes} \cr
\noalign{\horule}
\vsptwo u_s \; [\hbox{km}\, s^{-1}] && 153 & 165 & 132 && (1) \cr
\vsptwo \Tbn && 77^\circ\pm 7^\circ & 75^\circ\pm 7^\circ & 
   55^\circ\pm 16^\circ && (2) \cr
\vsptwo v_{th}(H^+) \; [\hbox{km}\, s^{-1}] 
   && 23\pm 2 & 29\pm 3 & 50\pm 2 && (3) \cr
\vsptwo v_{th}(He^{2+}) \; [\hbox{km}\, s^{-1}]
   && 26\pm 3 & \dots & \dots && (3,\, 4) \cr
\vsptwo n_{H^+} \; [\hbox{cm}^{-3}] 
   && 1.4\pm 0.1 & 0.37\pm 0.03 & 3.0\pm 0.3 && \cr
\vsptwo n_{He^{2+}} \; [\hbox{cm}^{-3}] 
   && 0.06\pm 0.005 & \dots & \dots && (4)\cr
\vsptwo B_1\; [10^{-5}\;\hbox{Gauss}] 
   && 2.05\pm 0.05 & 1.95\pm 0.05 & 4.30\pm 0.05 && (5) \cr
\vsptwo T_e\; [10^4\; K] && 6.4 & 6.5 & 12.4 && (6) \cr
\noalign{\horule}
\vsptwo T_{H^+}\; [K] && 3.2\times 10^4 & 4.9\times 10^4
   & 1.5\times 10^5 && (7) \cr
\vsptwo T_{He^{2+}}\; [K] && 1.6\times 10^5 & \dots & \dots && (4,\, 7) \cr
\noalign{\horule}
\vsptwo \machson && 4.21 & 4.17 & 2.15 && (8)\cr
\vsptwo \machalf && 4.05 & 2.39 & 2.44 && \cr
\vsptwo \machtot && 2.92 & 2.07 & 1.61 && (9)\cr
} \htrule }
}
{Notes: (1) Shock velocities (relative to solar wind)
for 91097 and 91147 were taken from Burton et al.
(1992), but that for 91118 was estimated from the value of \teq{\machson} 
inferred from the \teq{\machtot} given by Burton et al.  
(2) Shock obliquities were taken from Burton et al. (1992)
and their uncertainties are given in Balogh et al. (1995).  
(3) The thermal velocities \teq{v_{th}} are measured Gaussian widths.  
(4) Omissions in \teq{He^{2+}} data occur where such information was not
used in the present comparison.  
(5) Mean upstream fields \teq{B_1} are taken from magnetometer
measurements (Hoang et al. 1995). 
(6) Electron temperatures for 91097 and 91147 are from (Hoang et al. 1995),
while that for 91118 was estimated using the large scale averages of
Phillips et al. (1995).
(7) Ion temperatures are from \teq{T_i =m_i v_{th}^2/(2k)}.
(8) The \teq{\machson} values assume that \teq{n_e =n_{H^+}}, and have
included the dynamical effects of \teq{He^{2+}}.
(9) Here \teq{\machtot = \machson\machalf
/\sqrt{\machson^2+\machalf^2}}.
Uncertainties are quoted for measured quantities. 
See the text for further elaboration on these parameters.
}


\clearpage


\figureoutsmall{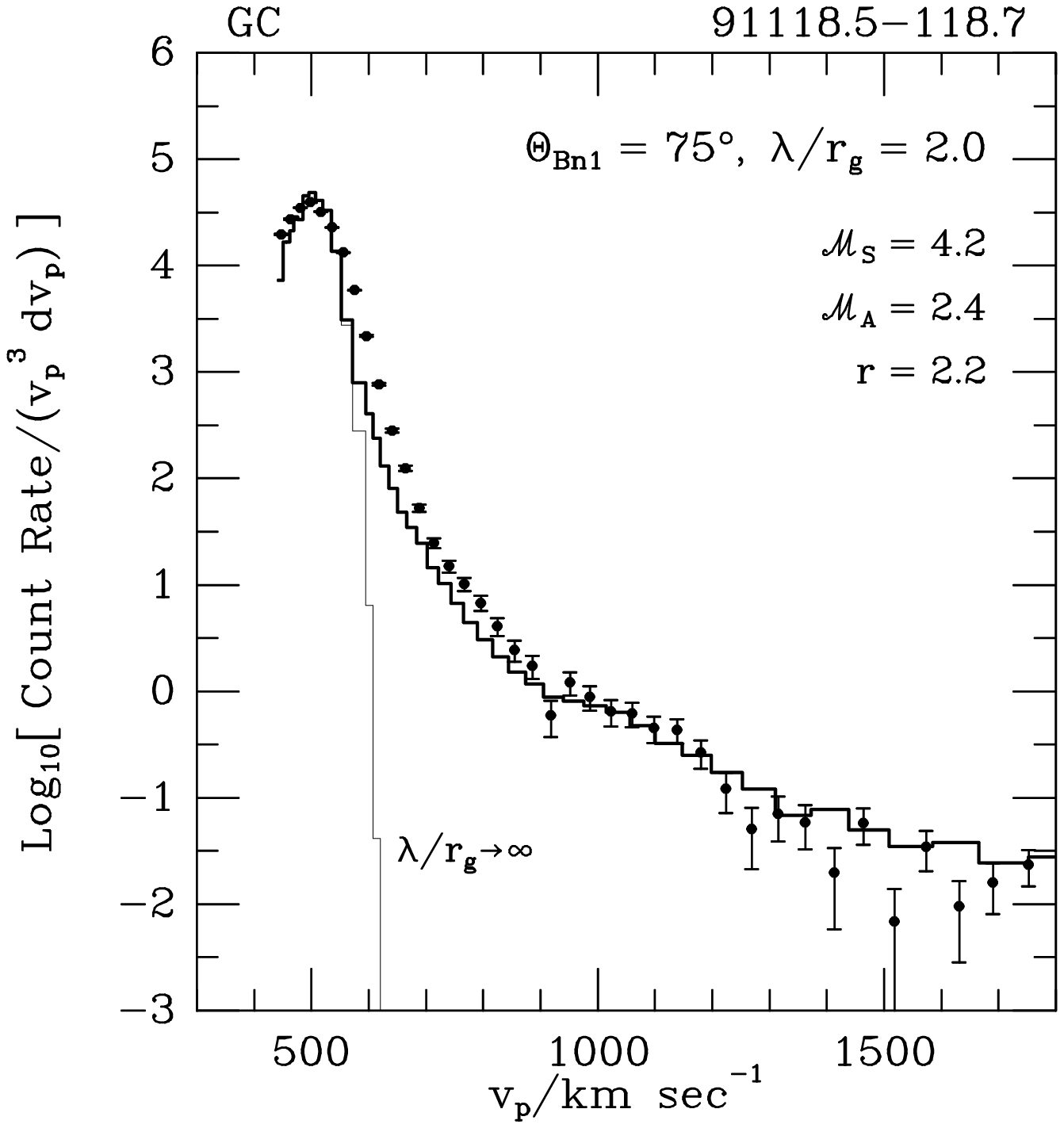}{
 Comparison of the SWICS \teq{H^+} data (points define the phase-space
 distribution, measured in the spacecraft frame and certain at the
 1\teq{\sigma} level) for the quasi-perpendicular interplanetary shock
 on April 28, 1991 (91118), taken over a period of a few hours on the
 downstream side where the distribution was relatively steady in time,
 and the Monte Carlo simulation output spectrum (histogram) from the
 guiding-center (GC) implementation of the code.  The mean measured
 shock obliquity of \teq{75^\circ} and other observed shock parameters
 (see Table~1) were used to yield model parameters such as the
 compression ratio, \teq{r}, and the sonic \teq{\machson} and
 Alfv\'enic \teq{\machalf} Mach numbers, as indicated.  The value of
 \teq{\lambda_0/r_{\rm g1}} (\teq{=\lambda /r_{\rm g}}), adjusted to
 obtain the fit, indicates that moderately strong scattering is
 operating in this shock.  The light solid histogram is the simulated
 distribution in the limit of \teq{\lambda/r_{\rm g}\to\infty}.  The
 count rate axis is expressed in units of km$^{-7}\;$sec$^4$.  For
 reference, proton speeds of \teq{1000}km/sec correspond to a kinetic
 energy of \teq{5.22}keV.       \label{fig:91118gc}
}

\figureout{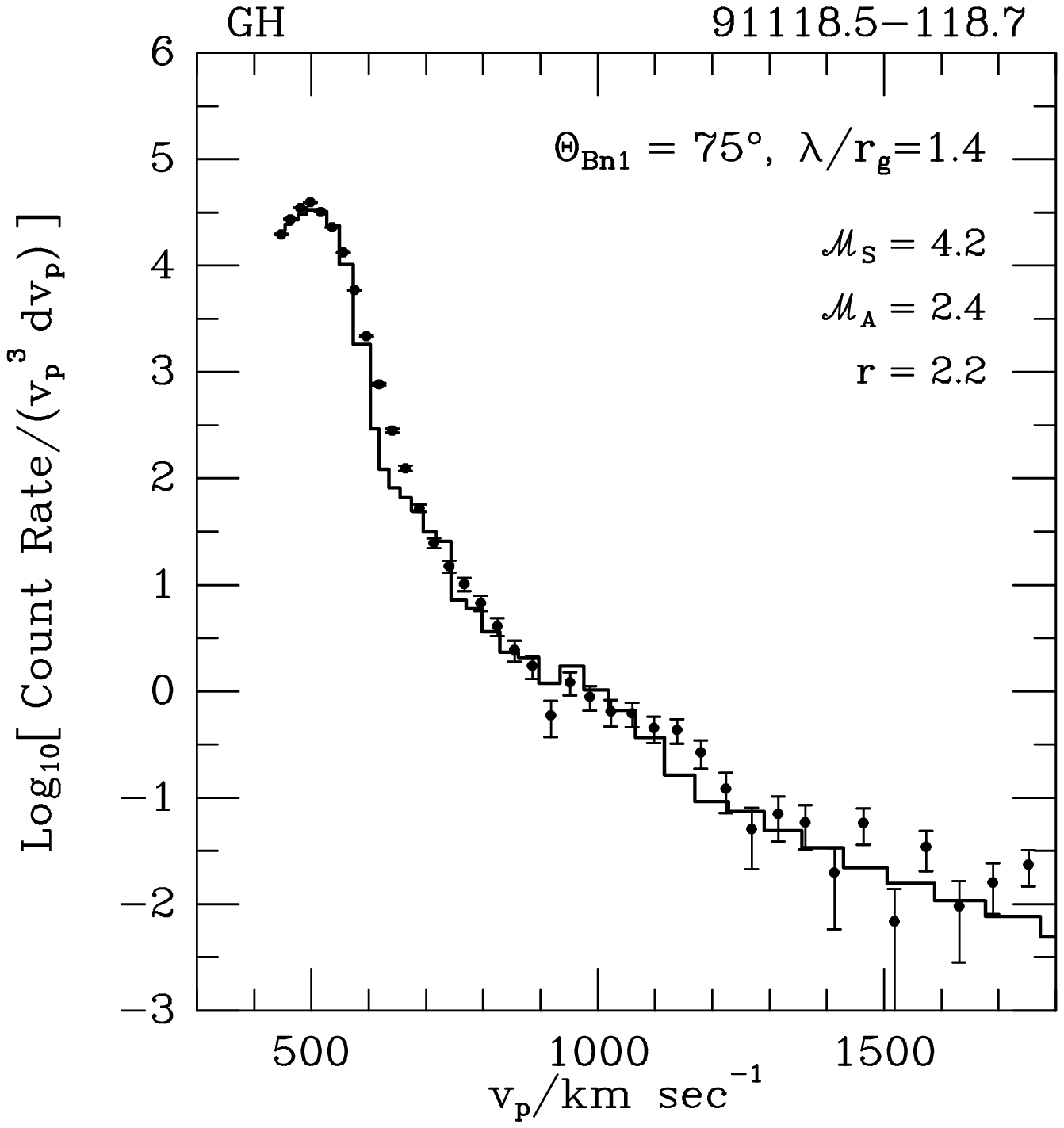}{
 The same SWICS proton data as in Fig.~1 are presented, but here are
 compared with an output spectrum from the version of the Monte Carlo
 code that computes particle gyrohelices (GH) exactly.  The similarity
 of the values of \teq{\lambda/r_{\rm g}} in Fig.~1 and obtained here that are
 required to fit the data indicate that the Monte Carlo model is largely
 insensitive to the details of particle gyromotions.  These extremely
 low values of \teq{\lambda/r_{\rm g}} are a direct consequence of the
 weakness of the shock (i.e low \teq{r}).             \label{fig:91118orb}
}

\figureout{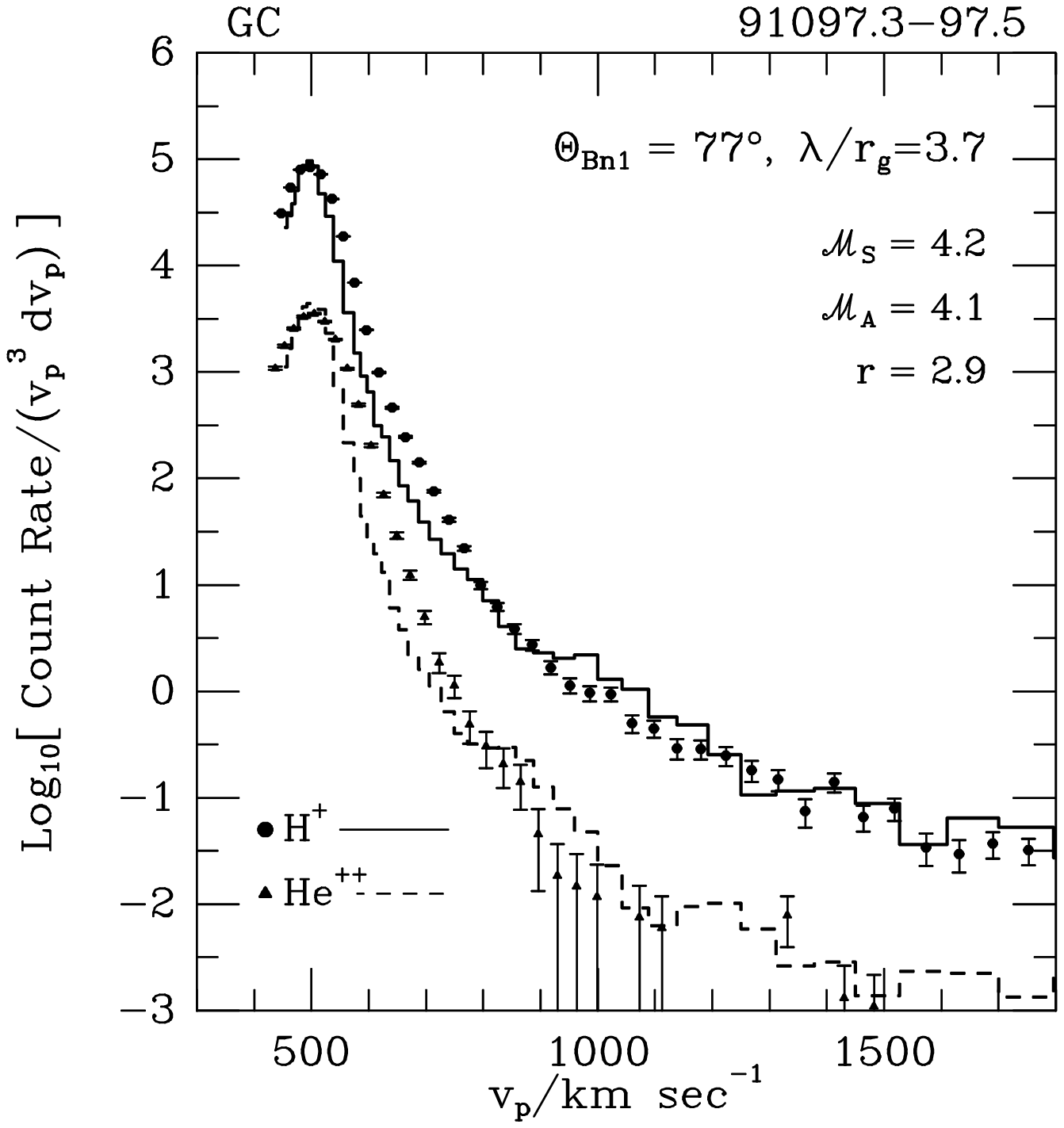}{
 Phase-space distributions for ionized hydrogen (filled circles) and 
 alpha particles (filled triangles), measured downstream of the April 7,
 1991 (91097) shock by SWICS, again over a period of several hours.  The
 solid and dashed histograms are the Monte Carlo simulation
 (guiding-center version, GC) predictions for \teq{H^+} and
 \teq{He^{2+}}, respectively, and the impressive fit implies the presence
 of very strong scattering, a species-independent acceleration process
 (see text for discussion), and also the absence of pick-up ions in this
 shock.  The higher compression ratio of this shock compared with 91118
 resulted in a higher fitting value of \teq{\lambda/r_{\rm g}}.   \label{fig:91097}
}

\figureout{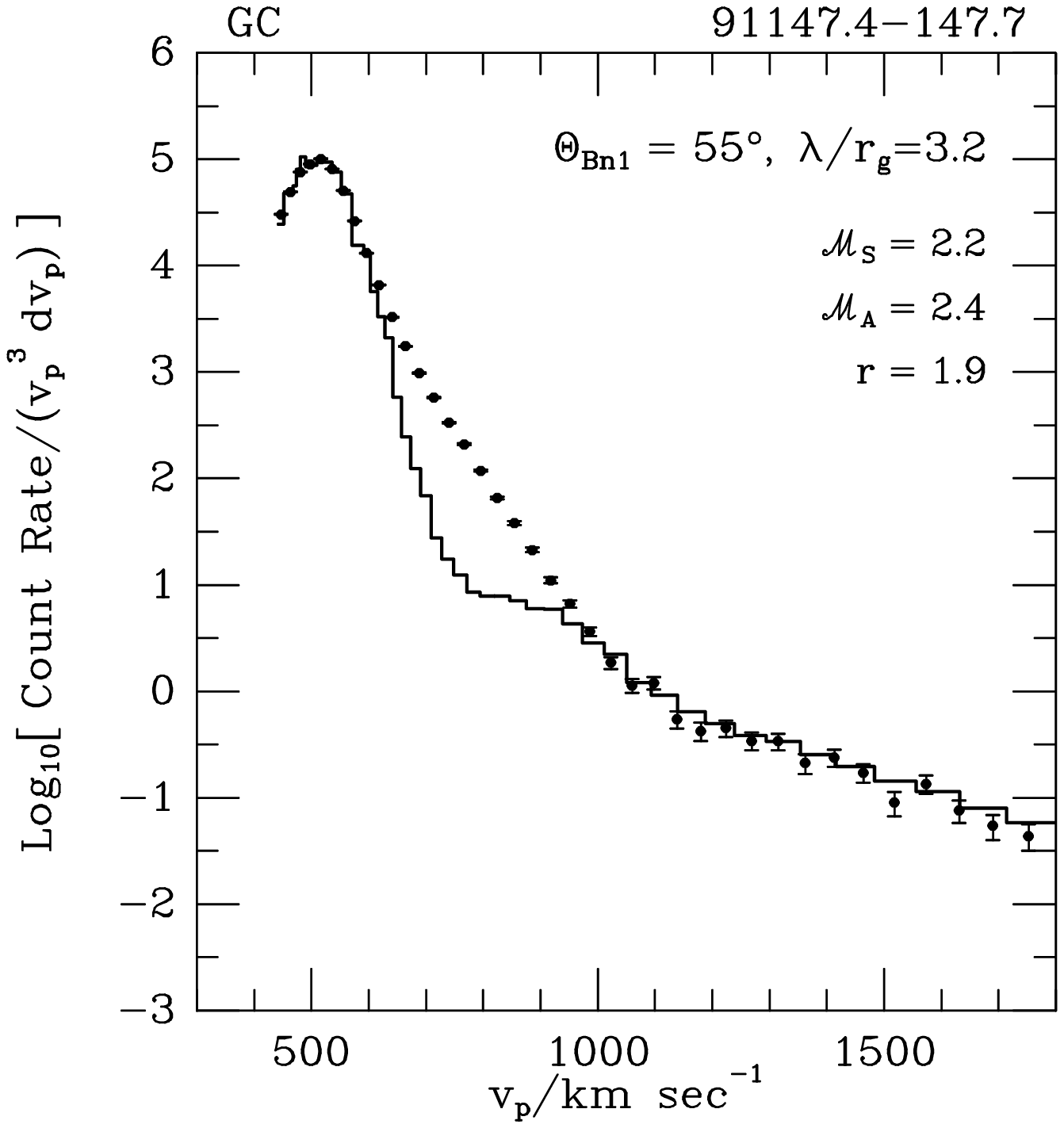}{
 A similar comparison to that in Figs.~1 and~3, but for the oblique
 interplanetary shock detected on May 27, 1991 (91147), again taken over
 a period of a few hours on the downstream side where the distribution
 was relatively steady in time.  The guiding-center version of the Monte
 Carlo code was again used to obtain the fit.  Strong scattering
 (\teq{\lambda /r_{\rm g} \lesssim 10}) is indicated.  The comparative
 inaccuracy of the fit in the suprathermal regime suggests the presence
 of another population of particles: the presence of pick-up ions might
 be inferred for this shock, which was about 0.45 AU further from the sun
 (see Table~1) than 91118.          \label{fig:91147}
}

\addtocounter{figure}{1}
\clearpage
%
%

\figureout{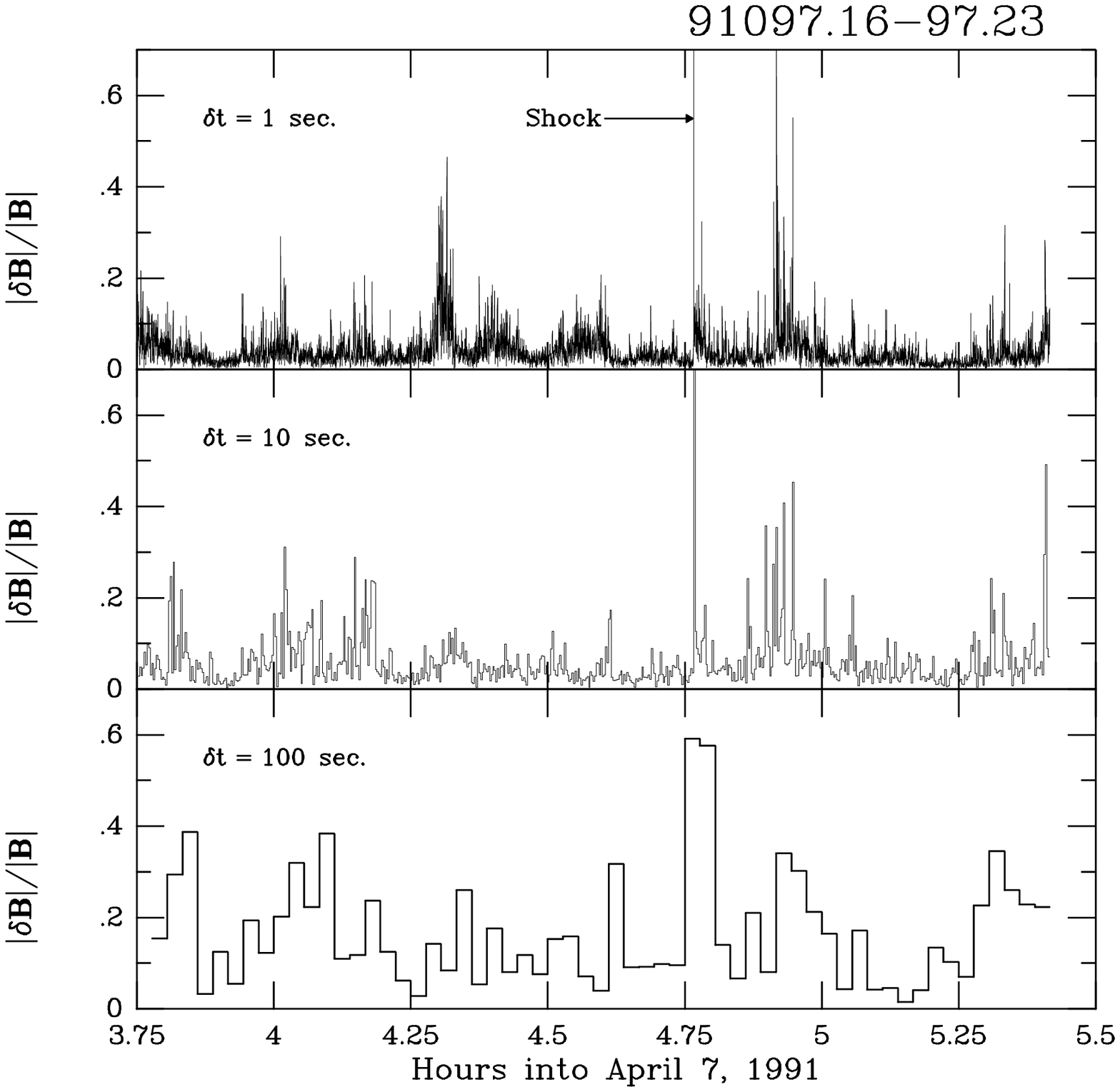}{
 The turbulence measure \teq{\vert \delta \hbox{\bf B}\vert/\vert
 \hbox{\bf B}\vert}, defined in Eq.~(\ref{eq:turb}), for times surrounding the
 91097 shock, obtained using the field data displayed in Fig.~5a. 
 Binning is exhibited on three timescales \teq{\delta t}, as labelled,
 indicating a slow increase in turbulence measure with timescale. 
 Clearly the degree of turbulence is similar either side of the shock,
 therefore indicating that in Fig.~5, \teq{\vert \delta \hbox{\bf
 B}\vert} scales with \teq{\vert \hbox{\bf B}\vert}.  The data at the
 shock for timescales of 1 and 10 seconds yield \teq{\vert \delta
 \hbox{\bf B}\vert/\vert \hbox{\bf B}\vert} in excess of unity. 
       \label{fig:turbulence}
}

\end{document}